\documentstyle[twocolumn,floats,epsf,ifthen,aps,prl]{revtex} 

\begin{document} 

\preprint{draft} 
 

\title{ Correlated quantum measurement of a solid-state qubit } 

\author{Alexander N. Korotkov}
\address{ 
Department of Physics and Astronomy, 
State University of New York, 
Stony Brook, NY 11794-3800
} 
\date{\today} 
 
\maketitle 
 
\begin{abstract} 
        We propose a solid-state experiment to study the process 
of continuous quantum measurement of a qubit state. 
The experiment would verify that an individual 
qubit stays coherent during the process of measurement (in contrast 
to the gradual decoherence of the ensemble-averaged density matrix) 
thus confirming the possibility of the qubit purification by 
continuous measurement. 
The experiment can be realized using quantum dots, single-electron
transistors, or SQUIDs. 
\end{abstract} 
 
\narrowtext 
 

        The impressive advantages promised by quantum computing  
\cite{Bennett} have revived the interest to the fundamental  
quantum effects in simple objects: two-level systems, which 
in this context are nowadays called qubits. 
In this paper we address the problem of continuous measurement 
of a qubit state having in mind a solid-state realization of the setup.

        Among the numerous proposals of quantum computers, the solid-state 
realizations (see, e.g.\ Refs.\ \cite{Kane,Averin,Makhlin,Mooij}) 
look more promising because of better controllability of qubit 
parameters and inter-qubit couplings. However, the qubit measurement 
in this case is not as straightforward as in typical optical experiments
where the single photon just ``clicks'' the detector. The reason 
is finite (and typically weak) coupling with a solid-state detector 
and finite intrinsic noise of the detector. As a result, the measurement 
cannot be done instantaneously, and so the collapse postulate of the 
``orthodox'' quantum mechanics \cite{Neumann} cannot be applied directly.
Instead, the quantum measurement should be considered as a continuous 
process. 

        There are two main theoretical approaches to the continuous quantum 
measurements. One approach (which dominates in solid-state physics 
and so can be called ``conventional'') is based on the theory of
interaction with dissipative environment \cite{Caldeira,Zurek}. 
Taking trace over the numerous degrees of freedom of the detector,
it is possible to obtain the gradual evolution of the density
matrix of the measured system from the pure initial state 
to the incoherent statistical mixture, 
thus describing the measurement process. Since the procedure implies
the averaging over the {\it ensemble}, the final equations of this 
formalism are deterministic and can be derived from the Schr\"odinger
equation alone, without any notion of the state collapse. 

        The other approach (see, e.g., Refs.\ 
\cite{Gisin,Carmichael,Gagen,Plenio,Mensky,Kor-PRB}) is closer to the 
collapse viewpoint and describes the stochastic evolution of an  
{\it individual} quantum system due to continuous measurement. 
This evolution obviously depends on a particular measurement 
result and is usually called selective or conditional quantum  
evolution. Depending on the details of the studied measurement setup 
and applied formalism, different authors 
\cite{Gisin,Carmichael,Gagen,Plenio,Mensky,Kor-PRB} 
discuss quantum trajectories, quantum state diffusion, stochastic evolution
of the wavefunction, quantum jumps, stochastic Schr\"odinger equation,
complex Hamiltonian, method of restricted path integral, Bayesian formalism, 
etc. 
The theory of selective quantum evolution was only recently introduced 
into the context of solid-state mesoscopics \cite{Kor-PRB,Kor-LT}. 
In particular, it was shown that the continuous 
measurement of an individual qubit does not lead to gradual decoherence 
(in contrast to the conventional result for the ensemble),  
instead, the measurement can lead to gradual purification of the 
qubit density matrix.

        Since the concept is still considered controversial, 
the experimental check is quite important. In this paper we propose 
an experiment which can be realized using three possible  
setups available for present-day technology: double-quantum-dot qubit 
measured by quantum point contact \cite{Buks}, 
qubit based on single-Cooper-pair box measured by single-electron 
transistor \cite{Nakamura}, or SQUID-based qubit  
measured by another SQUID \cite{Friedman,Mooij2}.  

\begin{figure} 
\centerline{
\epsfxsize=1.5in 
\hspace{0.5cm}
\epsfbox{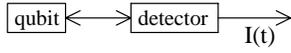}}  
\vspace{0.3cm} 
\caption{ 
Schematic of a qubit continuously measured by a detector with
output signal $I(t)$. 
 }
\label{schem1}\end{figure}

        Let us start with reviewing the result of the conventional formalism 
for the continuous measurement (Fig.\ \ref{schem1}) of a qubit state 
(see, e.g. recent  publications 
\cite{Buks,Gurvitz,Aleiner,Levinson,Stodolsky,Shnirman,Kor-Av}). 
For the qubit characterized by the standard Hamiltonian
${\cal H}_{QB} = (\varepsilon /2) (c_1^\dagger c_1 -c_2^\dagger c_2)
+H(c_1^\dagger c_2 +c_2^\dagger c_1)$ in the basis defined by 
coupling with the detector, the evolution of qubit density 
matrix $\rho$  is given by equations
        \begin{eqnarray}
&&\dot{\rho}_{11}=  -\dot{\rho}_{22}=  -2 H 
        \mbox{Im} \rho_{12}  , 
        \label{conv1}\\
&& {\dot\rho}_{12}=  \imath  \varepsilon \rho_{12}+ 
        \imath H  
(\rho_{11}-\rho_{22}) -\Gamma  \rho_{12} ,  
        \label{conv2}\end{eqnarray} 
where the continuous measurement is described by the dephasing rate $\Gamma$,
which was calculated for different models in Refs.\ 
\cite{Buks,Gurvitz,Aleiner,Levinson,Stodolsky,Shnirman,Kor-Av}. 

        These equations do not depend on the detector output because 
they represent the result of ensemble averaging, including 
the averaging over the measurement result. To study the evolution of 
an individual qubit let us denote the noisy detector signal as $I(t)$ 
(assuming current for definiteness). Two ``localized'' qubit states 
1 and 2 correspond to average detector currents $I_1$ and $I_2$
which by assumption do not differ much, $\Delta I \equiv I_1-I_2 \ll 
I_0 \equiv (I_1+I_2)/2$. (This assumption of ``weakly responding'' 
detector \cite{Kor-PRB} allows us to use the linear response theory  
and also Markov approximation if the processes in the detector are 
much faster than the qubit evolution.) 
Intrinsic noise of the detector signal is characterized by the
spectral density $S$ which is frequency-independent in the range 
of interest. The noise determines the typical measurement time 
$t \sim S/(\Delta I)^2$ necessary to distinguish between states 
1 and 2, and thus defines the timescale of the selective evolution
of the qubit density matrix $\rho (t)$. 

        Within the Bayesian formalism \cite{Kor-PRB} the selective 
evolution is described by equations 
        \begin{eqnarray}
\dot{\rho}_{11}= &&  -2H \mbox{Im} \rho_{12} 
 +(2\Delta I/S)\, \rho_{11}\rho_{22} [I(t)-I_0], 
        \label{Bayes1}\\
 {\dot\rho}_{12}= &&  \imath \varepsilon \rho_{12} 
        + \imath H (\rho_{11}-\rho_{22}) 
\nonumber \\ 
&&  - (\Delta I/S) \, ( \rho_{11}-  \rho_{22})  
[I(t)-I_0] \, \rho_{12} -\gamma \, \rho_{12},  
        \label{Bayes2} \end{eqnarray} 
where the dephasing $\gamma =\Gamma - (\Delta I)^2/4S \geq 0$  
is now due to the contribution from ``pure environment'' only. 
In particular,
$\gamma =0$ if the qubit is measured by symmetric quantum point
contact, since in this case $\Gamma =(\Delta I)^2/4S$ 
(see Refs.\ \cite{Gurvitz,Aleiner,Buks,Kor-Av}). We will call
such detector an ideal detector, $\eta =1$, where 
$\eta \equiv 1-\gamma / \Gamma$ is the ideality factor. 
 In contrast, the single-electron transistor \cite{Av-Likh} in the operation 
point far outside the Coulomb blockade range  is a significantly 
nonideal detector \cite{Shnirman}, $\eta \ll 1$; however, $\eta$ becomes
comparable to unity when the current is mostly due to cotunneling 
processes \cite{Kor-sp}. 
        The SQUID is an ideal detector when its sensitivity is 
quantum-limited \cite{Danilov,Av-sp}.

        Eqs.\ (\ref{Bayes1})--(\ref{Bayes2}) allow us to calculate
the evolution of qubit density matrix $\rho$ if the detector output $I(t)$ 
is known from a particular experiment. To simulate the measurement 
we can use the replacement \cite{Kor-PRB}
                \begin{equation}
I(t)- I_0 \, = \, \Delta I (\rho_{11}-\rho_{22})/2 +\xi (t) , 
        \label{Bayes3}\end{equation}
where the random process $\xi (t)$ has zero average and ``white'' spectral
density $S_\xi =S$. One can check that averaging of Eqs.\ 
(\ref{Bayes1})--(\ref{Bayes2}) 
over all possible measurement results [i.e.\ over random contribution  
$\xi (t)$] reduces them to Eqs.\ (\ref{conv1})--(\ref{conv2}). 
Notice that the stochastic equations are written 
in Stratonovich form which preserves the usual calculus rules, while 
averaging is more straightforward in It\^o form \cite{Oksendal}. 

        As follows from Eqs.\ (\ref{Bayes1})--(\ref{Bayes2}),
if a qubit with initially pure state, $|\rho_{12}(0)|^2=
\rho_{11}(0)\rho_{22}(0)$, is measured by an ideal detector, then 
its density matrix $\rho (t)$ stays pure during the measurement process. 
Even if initial state is a statistical mixture, $\rho (t)$ is gradually 
purified during the measurement \cite{Kor-PRB}. 

        The predictions of the Bayesian formalism 
can be checked  
experimentally, however, it is not quite simple at the present-day level of 
solid-state technology. The direct experiment was discussed in Ref.\ 
\cite{Kor-PRB}. The idea was to perform the measurement by almost ideal 
detector during some finite time $\tau$, record the detector output $I(t)$, 
use Eqs.\ (\ref{Bayes1})--(\ref{Bayes2}) to calculate $\rho (\tau )$ 
and then check the calculated value. This check can be done by changing 
qubit parameters $\varepsilon $ and $H$ in a way to ensure $\rho_{11}=1$ 
at some specified moment of time, that can be measured by the detector
switched on again. Since for coherent evolution the qubit can be placed
with 100\% certainty in the state 1 only if the wavefunction is known  
precisely, such check (repeated many times) verifies that $\rho (\tau )$
is pure and coincides with the calculated value. 

        Unfortunately, this experiment would require very fast recording
of $I(t)$. Since the expected coherence time is on the order of 10--100 ns 
at most (see, e.g.\ \cite{Nakamura}), the bandwidth of  the detector signal 
coming
out of the cryostat should be at least 1 GHz, that is very difficult 
experimentally. Another proposed experiment \cite{Kor-Av,Kor-sp} is
to measure the spectral density of the quantum coherent oscillations
and check the predicted maximal peak-to-pedestal ratio of 4. 
Such an experiment may be easier to realize (because the basic spectral 
analysis can be done on-chip inside the cryostat), however, it would 
not prove unambiguously the Bayesian formalism, since an alternative 
interpretation of the result is possible \cite{Kor-sp}.

\begin{figure} 
\centerline{
\epsfxsize=2.5in 
\hspace{0.3cm}
\epsfbox{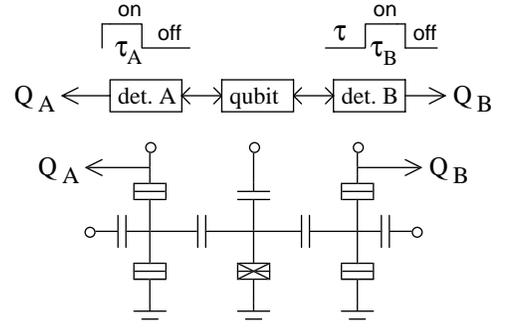}
        }  
\vspace{0.3cm} 
\caption{ 
Schematic of two-detector correlation experiment using
Cooper-pair box and two single-electron transistors.
 }
\label{schem2}\end{figure}

        Here we propose an experiment which is even easier
to realize, and which can test the Bayesian formalism 
(\ref{Bayes1})--(\ref{Bayes2}). The main idea is to use two detectors 
($A$ and $B$) connected to the same qubit (Fig.\ \ref{schem2}). 
The detectors are switched on for short periods of time 
by two shifted in time voltage pulses (one for each detector) 
with durations $\tau_A$ and $\tau_B$, supplied 
from the outside. The output signal from the detector $A$ is the 
total charge $Q_A=\int_0^{\tau_A} I_A(t)\, dt$ passed during the measurement 
period. Similarly, the output from the detector $B$ is 
$Q_B=\int_{\tau}^{\tau +\tau_B} I_B(t)\, dt$, where $\tau$ is the time shift 
between pulses. If the measurement by the detector $A$ changes the qubit
density matrix, it will affect the result of measurement $B$. Repeating
the experiment many times (with the same initial qubit state) 
we can obtain the probability distribution  
$P(Q_A, Q_B |\tau )$ of different outcomes, which contains the information
about the effect of the quantum measurement on the qubit density matrix. 
        In comparison with previous suggestions, the advantage of 
this correlation experiment is that the wide signal bandwidth 
is required only for input pulses (that is relatively simple) 
while the outputs are essentially low frequency signals. 
The experiment can be called ``Bell-type'' because of some 
similarity with the famous proposal of Ref.\ \cite{Bell}. 

        Fig.\ \ref{schem2} shows the realization of the experiment 
using single-electron transistors (two small tunnel 
junctions in series \cite{Av-Likh}) as detectors. Qubit is realized
by the Cooper-pair box \cite{Lafarge,Nakamura} so that the electric charge
of the central island can be in coherent combination of two discrete 
charge states. Another similar setup is two quantum point contacts 
measuring the charge state of a double-quantum-dot qubit. One more
setup is the 3-SQUID experiment in which the qubit is realized by 
one SQUID while two other SQUIDs are in the detecting regime. 
For definiteness we will consider the realization of Fig.\ \ref{schem2}. 

        The conventional formalism (\ref{conv1})--(\ref{conv2})
does not give any explicit predictions for the resulting probability 
distribution $P(Q_A, Q_B |\tau)$. However, it implies the absence
of correlations between $\rho (t)$ and $I(t)$, so for example the average 
result of the second measurement $\overline{Q}_B(Q_A,\tau)\equiv
\int Q_B P(Q_A, Q_B |\tau)\, dQ_B$ should not depend on $Q_A$. The Bayesian
formalism (\ref{Bayes1})--(\ref{Bayes2}) makes the different prediction: 
$\overline{Q}_B$ does depend on $Q_A$.

        For simplicity let us assume symmetric qubit, $\varepsilon =0$, 
which is initially in 
the ground state, $\rho_{11}=\rho_{22}=\rho_{12}=0.5$, and also assume 
relatively strong coupling between the qubit and detectors, 
$(\Delta I_A)^2/HS_A \gg 1$, $(\Delta I_B)^2/HS_B \gg 1$ (subscripts
$A$ and $B$ correspond to two detectors), so that we can neglect
the qubit evolution due to finite $H$ during the measurement
periods $\tau_A$ and $\tau_B$, which are assumed to be on the order
of $S_{A,B}/(\Delta I_{A,B})^2$. Then from Eqs.\ 
(\ref{Bayes1})-({\ref{Bayes2})
if follows that the first measurement only ``partially'' localizes 
the qubit state and 
after obtaining the result $Q_A$ from the first measurement 
the qubit density matrix is
        \begin{eqnarray}
&& 2 \rho_{11}(\tau_A)-1 =  
\tanh \frac{(Q_A-\tau_A I_{2A})^2- (Q_A-\tau_A I_{1A})^2} 
{2S_A \tau_A} \, , 
        \label{ta1}\\ 
&& \rho_{12}(\tau_A)=[\rho_{11}(\tau_A) \, \rho_{22}(\tau_A)]^{1/2} 
\exp(-\gamma_A \tau_A) \, , 
        \label{ta2}\end{eqnarray}
where Eq.\ (\ref{ta1}) is actually the classical Bayes formula which was
used in Ref.\ \cite{Kor-PRB} to derive the formalism 
(\ref{Bayes1})--(\ref{Bayes2}).
[The probability to get $Q_A$ has the distribution 
$P(Q_A)=(p_1+p_2)/2$ where $p_i=(\pi S_A\tau_A)^{-1/2} 
\exp (-(Q_A-\tau_AI_{iA})^2/S_A\tau_A)$.]  
 The qubit performs the free evolution 
during the time $\tau -\tau_A$ between measurements (here we neglect 
$\tau_A\ll \tau$) and the average result of the second 
measurement $\overline{Q}_B=\tau_B (I_{2B}+ \rho_{11}(\tau )\Delta I_B)$ 
depends on $Q_A$ in the following way (Fig.\ \ref{fig3}a): 
        \begin{eqnarray} 
\delta_B =&&  \frac{1}{2} \tanh \frac{(Q_A-\tau_A I_{2A})^2-
(Q_A-\tau_A I_{1A})^2} 
{2S_A \tau_A} 
        \nonumber \\
&& \times \frac{2H}{\Omega} \cos \left( \Omega \tau -
\arcsin \frac{\gamma_f}{4H}\right) \exp (-\gamma_f \tau /2 ) \, , 
        \label{aver1}\end{eqnarray} 
where $\delta_B\equiv (\overline{Q}_B-\tau_B I_{0B})/\tau_B\Delta I_B$, 
$\gamma_f$ 
is the dephasing with both detectors switched off, and 
$\Omega =(4H^2-\gamma_f^2/4)^{1/2}$ is the frequency of quantum oscillations
(underdamped case is assumed). Notice that $\delta_B$ changes sign 
together with the sign of $Q_A-\tau_A I_{0A}$, while the phase of 
oscillations is a piece-constant function of $Q_A$. 

\begin{figure} 
\centerline{
\epsfxsize=3.1in 
\vspace{0.3cm}
\hspace{0.4cm}
\epsfbox{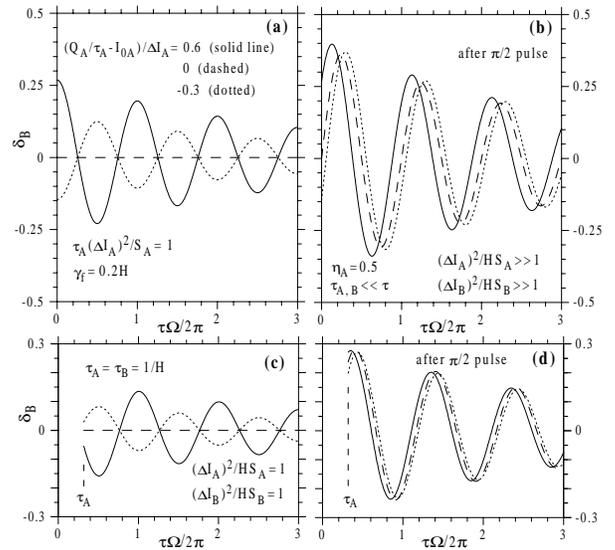} 
}  
\vspace{0.1cm} 
\caption{ The normalized average result $\delta_B$ of the second measurement 
for several selected results $Q_A$ of the first measurement, 
as a function of the time $\tau$ between measurements. 
Panels (a)--(b) are for strong coupling and panels (c)--(d) for 
moderate coupling between the qubit and detectors (other parameters 
are the same). 
The calculations are done by Bayesian formalism while the conventional 
formalism does not predict any nontrivial dependence. 
 }
\label{fig3}\end{figure}

        The dependence becomes quite different if the $\pi /2$ pulse is 
applied to the qubit immediately after the first measurement, that 
multiplies $\rho_{12}(\tau_A)$ given by Eq.\ (\ref{ta2}) by the 
imaginary unit. In this case (Fig.\ \ref{fig3}b) 
        \begin{eqnarray}
\delta_B && = A \sin (\Omega \tau +\arcsin z/A) \exp (-\gamma_f\tau /2)\, , 
        \nonumber \\
&& A=  
  [ (z^2+y^2-yz\gamma_f/2H)/
(1-\gamma_f^2/16H^2) ]^{1/2} ,  
        \label{aver2}\end{eqnarray}
where $z=\rho_{11}(\tau_A)-1/2$ and $y=\mbox{Im}\rho_{12} (\tau_A +0)=
\mbox{Re}\rho_{12}(\tau_A -0)$ are
given by Eqs.\ (\ref{ta1})--(\ref{ta2}). This expression considerably
simplifies for weak dephasing, $\gamma_A\tau_A \ll 1$ and $\gamma_f \ll H$,
when 
\begin{equation}
 \delta_B    = \frac{1}{2} \sin [ \Omega \tau +
\arcsin (2\rho_{11}(\tau_A)-1)]  
 \exp (-\frac{\gamma_f\tau}{2}).
\label{aver3}\end{equation}
	  

        In contrast to Eq.\ (\ref{aver1}) now the phase of 
oscillations $\delta_B(\tau )$ depends on the result $Q_A$ of 
the first measurement, while the amplitude is maximal possible and 
independent of $Q_A$. This fact is very important since it {\it proves} that
after the first measurement (by an ideal detector) the qubit remains 
in the pure state for {\it any} result $Q_A$. This state depends on 
$Q_A$ and is not one of the localized states as somebody could naively
expect. [Notice that Eq.\ (\ref{aver1}) can in principle be interpreted 
in terms of such ``classical'' localization, as indicated by its
independence on $\eta_A$.] 

        In a realistic experimental situation the assumption 
of strong coupling with detectors may be inapplicable. 
In this case the full probability distribution $P(Q_A, Q_B | \tau)$
as well as the dependence $\overline{Q}_B(Q_A,\tau)$ can be calculated
numerically using Eqs.\ (\ref{Bayes1})--(\ref{Bayes3}).
The results of these calculations for $(\Delta I_A)^2/HS_A=
(\Delta I_B)^2/HS_B=1$ are shown in Figs.\ \ref{fig3}c and \ref{fig3}d.
Weak coupling as well as the nonideality of the detectors decrease  
the correlation between the results of two measurements, however,
for moderate values of the coupling and nonideality the correlation 
is still significant. 

        The successful experimental demonstration of the correlation 
and quantitative agreement with the results of the Bayesian formalism 
would prove the validity of this formalism and therefore prove its 
other predictions. 
	Besides the clarification of the relation between the measurement
result and qubit evolution, 
the important for practice prediction is the gradual qubit
purification  due to continuous measurement 
which can be useful for a quantum computer. 

        All quantum algorithms require the supply of ``fresh'' qubits
with well-defined initial states. This supply is not a trivial problem since 
the qubit left alone for some time deteriorates due to interaction 
with environment. The usual idea is to use the ground state which
should be eventually reached and does not deteriorate. However, 
to speed up the qubit initialization we need to increase the coupling 
with environment that should be avoided. The other possible idea 
is to perform the projective measurement after which the state 
becomes well-defined. However, in the realistic case the coupling
with the detector is finite that makes projective measurements impossible. 
Here we propose a different way: to tune qubit continuously in order 
to overcome the dephasing due to environment and so keep qubit ``fresh''.

\begin{figure} 
\centerline{
\epsfxsize=2.5in 
\hspace{0.3cm}
\epsfbox{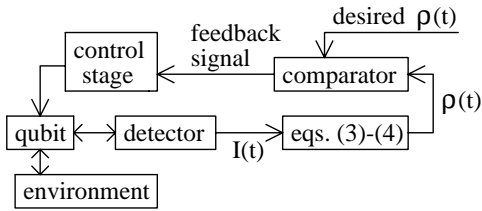}
        }  
\vspace{0.3cm} 
\caption{ 
Schematic of the continuous qubit purification using the quantum 
feedback loop. 
 }
\label{purification}\end{figure}

        The idea of such state purification is shown in Fig.\ 
\ref{purification}. The qubit 
is continuously measured by weakly coupled detector, and the detector 
signal is plugged into Eqs.\ (\ref{Bayes1})--(\ref{Bayes2}) which
allow us to monitor the evolution (in particular, due to interaction 
with environment) of qubit density matrix $\rho_{ij}(t)$. This evolution 
is compared with the desired evolution and the difference is
used to generate the feedback signal which controls the qubit
parameters $H$ and $\varepsilon$ in order to reduce the difference with
the desired qubit state. We have performed the Monte-Carlo simulation 
of the qubit purification by feedback loop (in the regime of well-pronounced
quantum oscillations) and found strong suppression 
of the qubit dephasing due to environment in the case when the dephasing
rate $\gamma$ 
is comparable or weaker than the ``measurement rate'' $(\Delta I)^2/4S$.

        In conclusion, we have proposed the Bell-type experiment which 
can test the Bayesian formalism predictions for the evolution of an 
individual qubit due to continuous quantum measurement. The next (and much 
more difficult) step is the experimental realization of the
qubit purification using quantum feedback loop.  

        The author thanks L. P. Rokhinson, D. V. Averin, M. H. Devoret, 
C. M. Marcus, and K. K. Likharev 
for useful discussions.

\end{document}